\documentclass{aa}
\usepackage{graphicx}
\usepackage{natbib}
\usepackage{longtable}

\newcommand{\pri}    {${\rlap.}^{\prime \prime}$}

\newcommand{\ltsima} {$\; \buildrel < \over \sim \;$}
\newcommand{\simlt}  {\lower.5ex\hbox{\ltsima}}            
\newcommand{\gtsima} {$\; \buildrel > \over \sim \;$}
\newcommand{\simgt}  {\lower.5ex\hbox{\gtsima}}            

\usepackage{txfonts}
\begin{document}

\title{A new search strategy for microquasar candidates using NVSS/2MASS and {\it XMM-Newton} data}


\author{J.~A. Combi\inst{1}, J.F. Albacete-Colombo\inst{2,3}, J. Mart\'{\i}\inst{1}
          }
\authorrunning{Combi et~al.}
\titlerunning{New microquasars candidates in the Galaxy}
\offprints{J.A. Combi}

\institute{Departamento de F\'{\i}sica (EPS), Universidad de Ja\'en,
Campus Las Lagunillas s$/$n, Ed-A3, (23071) Ja\'en, Spain\\
\email{jcombi@ujaen.es, jmarti@ujaen.es}
\and
Centro Universitario Regional Zona Atl\'antica (CURZA). Universidad Nacional del COMAHUE, Monse\~nor Esandi y Ayacucho (8500), Viedma (Rio Negro), Argentina. 
\email{albacete@fcaglp.unlp.edu.ar}
\and
Osservatorio Astronomico di Palermo, Piazza del Parlamento 1, Palermo (90141), Italy\\
             }

\date{Received; accepted }


  \abstract
   {}
   {Microquasars are ideal natural laboratories for understanding accretion/ejection
   processes, studying the physics of relativistic jets, and testing gravitational
   phenomena. Nevertheless, these objects are difficult to find in our Galaxy.
   The main goal of this work is to increase the number of known systems of this kind, which should allow
   better testing of high-energy phenomena and more realistic statistical studies of this galactic population to be made.}
   {We have developed an improved search strategy based on positional cross-identification 
   with very restrictive selection criteria to find new MQs, taking advantage of more sensitive modern 
   X-ray data. To do this, we made combined use of the radio, infrared, and X-ray
   properties of the sources, using different available catalogs.} 
  {We find 86 sources with positional coincidence in the NVSS/XMM catalogs at
galactic latitudes $|$b$|$ $\leq$ 10$^{\circ}$. Among them, 24 are well-known objects 
and the remaining 62 sources are unidentified. Out of 86 sources, 31 have one or two possible
infrared counterparts in the 2MASS catalog. For the fully coincident sources, whenever possible, 
we analyzed color-color and hardness ratio diagrams and found that at least 3 of them display 
high-mass X-ray binary characteristics, making them potential microquasar candidates.}
{}


\maketitle
%

\section{Introduction}

Some of the most attractive objects in the Galaxy are the
enigmatic microquasars (MQs). On a smaller scale, they copy the
characteristics exhibited by distant quasars \cite{mirabel99}. These
systems are X-ray binaries (XRBs) containing compact objects like
stellar black holes or neutron stars that accrete matter from a
companion star. They are known to emit from radio to X-ray energies
\cite{mirabel94} and possibly up to TeV gamma-ray energies, as in the
case of Cygnus X-1 \cite{albert07}. Evidence that jets of accreting X-ray binaries can
accelerate particles to TeV  energies has nevertheless been presented by Corbel et
al. (2002). MQs combine two important aspects of relativistic astrophysics: accreting black 
holes or neutron stars identified by the
production of hard X-rays around accreting disks and
relativistic radio jets detected by means of their synchrotron
emission.

These binary systems are ideal natural laboratories for understanding 
accretion/ejection process and other gravitational phenomena. However,
they seem to be rare objects in our Galaxy. In order to enable more
robust statistical studies, it is necessary to increase the number of
known MQs. Of the 15 currently confirmed MQs in the galaxy, 6 belong
to the high-mass X-ray binary (HMXB) class and 9 are of the low-mass
X-ray binary (LMXB) kind \cite{paredes05}.

Finding new MQs candidates is not an easy task. Considerable effort 
in the past has been put in to increasing their number. A few
searches for radio-emitting XRB systems and therefore new MQ
candidates, based on cross-identifications between radio, infrared, and
X-ray catalogs, have been carried out in the past, but without enough
success \citep[e.g.][]{paredes02,ribo04}. The method of looking for
such objects in the Galaxy usually includes a number
of steps with very restrictive selection criteria for the sources
being investigated. Mainly, a number of competing emission mechanisms
and several physical parameters should be associated to the same source
or system, and they basically include the properties of jet emission and
XRB behavior. The detection of emission at radio wavelengths could be
the signature that such  relativistic jets, mainly emitting via
incoherent synchrotron emission from very high-energy electrons
spiraling in magnetic fields, are present in the object. Ultraviolet
and infrared emissions are characteristics displayed by the normal star companions, and
X-ray radiation is most efficient at revealing accretion-powered
sources, such as binary stars.

As an example, a previous list of MQ candidates obtained from
cross-correlation of the radio NVSS catalog \cite{condon98} and
the ROSAT Bright Source Catalog (RBSC) sources of the Galactic plane
was presented by Paredes et al. (2002). They found 35 possible MQ
candidates based on a hardness ratio criterion  i.e.
$HR1+\sigma(HR1)\geq 0.9$, where $HR1=([0.5-2.0 {\rm ~keV}]-[0.1-0.4
{\rm ~keV}])/([0.5-2.0 {\rm ~keV}]+[0.1-0.4 {\rm ~keV}])$. However,
since ROSAT soft X-rays are strongly absorbed by the interstellar
medium, their resulting list of MQ candidates could be fairly reduced.
Given the impossibility that ROSAT will detect X-ray photons at energies
above 2.4 keV, and because X-rays in MQs are expected to be highly
energetic, such a criterion could not be efficient enough to select 
possible MQ candidates.

Here, we develop an improved search strategy that is also based on very
restrictive, but improved, selection criteria aimed at finding new MQs in
the Galaxy. Moreover, we take advantage of modern and
more sensitive multiwavelength data. In Sect.~\ref{search} we describe the search
strategy and sample definition. The general analysis and statistical
results are presented in Sect.~\ref{results}. Finally, we summarize
our main conclusions in Sect.~\ref{summary}.

\section{Search strategy and sample definition} \label{search}

We performed a positional cross-identification of the NRAO VLA
Sky Survey (NVSS\footnote{http://http://www.cv.nrao.edu/nvss/})
catalog \cite{condon98} with the Second {\it XMM-Newton} Serendipitous
Source Catalogue
(SXSSC\footnote{http://xmm.vilspa.esa.es/external/xmm$_{-}$data$_{-}$acc/xsa/index.shtml}).
The NVSS was observed with the array in D-configuration (DnC
configuration for the southernmost fields), which provides an
angular resolution of $45^{\prime\prime}$. It covers the entire sky from
the north pole to declination $-40^{\circ}$, and it contains $>$
1.8$\times$10$^{6}$ sources (296846 with $|b| \leq 10^{\circ}$),  with
a mean source number density of 51.8 per square degree and a
source density near the Galactic plane (for $|b| \leq 5^{\circ}$) of
82.4 per square degree. The rms positional uncertainties in RA
and DEC in the NVSS catalog  vary from $1^{\prime\prime}$ for
relatively strong (S$\geq$15 mJy) point sources to $7^{\prime\prime}$
for the faintest (S$\leq$2.3 mJy) detectable sources. Median positional uncertainties 
for sources with radio fluxes greater and lower than 15 mJy are, 1\pri6 and 4\pri8,
respectively. The completeness limit is about 2.5 mJy.

At the X-ray wavelengths, the SXSSC contains 153105 sources for which
24672 sources have $|b|$ $\leq$ 10$^{\circ}$. In this case the
mean source number density, for $|b| \leq 5^{\circ}$, is 6.85 per
square degree. This is a lower value than for NVSS radio sources because
the current XMM catalog only covers $\sim$ 2\% of the 
Galactic plane for $|b| \leq 10^{\circ}$. The systematic 1
sigma error on the SXSSC detection position ranges between 0\pri5
and about 1\pri5. In addition to the X-ray flux of source, RA and DEC
errors essentially depend on the source off-angle with respect to the
pointing center of the observation from which it was extracted\footnote{Details
about determining the positional source error can be follow at:
http://xmmssc-www.star.le.ac.uk/Catalogue/}. The median flux (in the
total photon energy band 0.2 - 12 keV) of the catalog sources is
$\sim$2.4 $\times$ 10$^{-14}$ erg cm$^{-2}$ s$^{-1}$; $\sim$ 20\% have
fluxes below 1$\times$ 10$^{-14}$ erg cm$^{-2}$ s$^{-1}$. The use of
XMM data is one of the main improvements of this paper over 
similar, previous works based on ROSAT data \citep{paredes02,ribo04}.

Since our intention is to find new galactic MQ candidates in the
Galaxy, we adopted a set of selection criteria for NVSS/SXSSC cross-identification 
summarized as follows:

\begin{enumerate}

\item Sources with galactic latitude $\leq$ 10$^{\circ}$ have been
selected from the NVSS and XMM catalogs. This significantly reduces 
the number of sources to about 16\% of the total for both catalogs.

\item No extended radio sources are included in the selection process.

\item We  assume that the positional coincidence has high significance when
the offset between NVSS/XMM is strictly R$\leq$ ($\epsilon_{\rm
XMM}$+$\epsilon_{\rm NVSS}$), where $\epsilon_{\rm XMM}$ and
$\epsilon_{\rm NVSS}$ are the absolute (RA,DEC) error in the position
of the XMM and NVSS sources, respectively. For this purpose, we used 
our IDL (Interactive Data Language\footnote{http://physics.nyu.edu/grierlab/idl$_{-}$html$_{-}$help/home.html}) 
based code to cross-correlate the aforementioned catalogs.

\item After that, we inspected the SIMBAD database and the NASA/IPAC
Extragalactic Database (NED) for detecting those previously known
sources and well-known MQs in the sample.

\item We further computed the hardness ratio of the whole sample and 
selected sources with hard X-ray spectra characterized by a power-law model with 
index $\Gamma \leq 2.5$ typical of LMXB and HMXB systems
\cite[e.g.][]{white95,sugi01}.

\item The positions of those NVSS/XMM coincident sources were
filtered afterwards with the 2MASS\footnote{http://irsa.ipac.caltech.edu/cgi-bin/Gator/} catalog
\cite{cutri03}, adopting a cross-identification radius of 4 arcsec, and we used a
color-color diagram $(J-H)/(H-K)$ to further constrain our search to
likely Galactic objects (see Sec.\,3.3 for more details).

\end{enumerate}

\begin{figure}
\center \resizebox{\hsize}{!}{\includegraphics[angle=0]{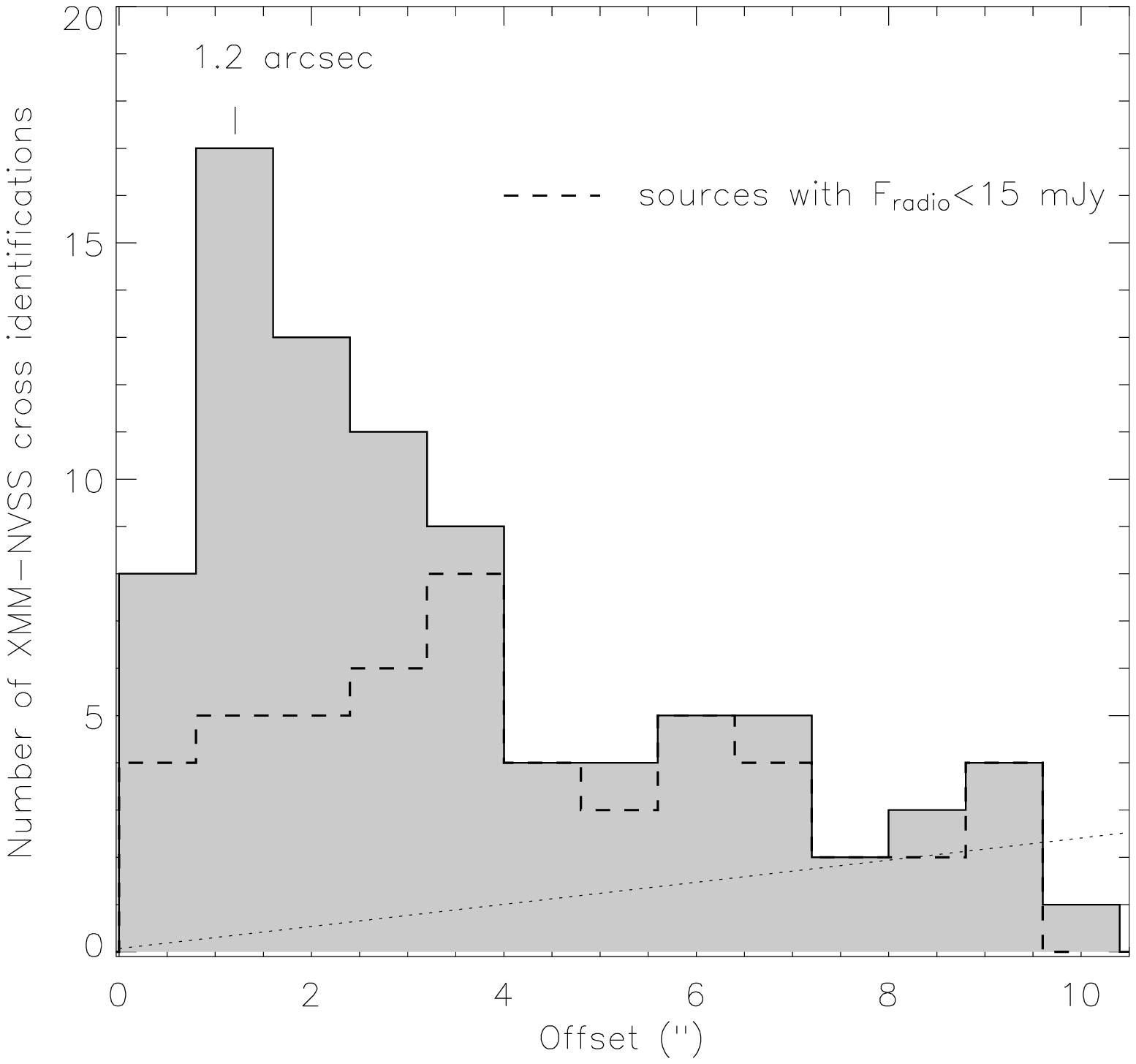}}
\caption{Numerical distribution of the positionally coincident
NVSS/XMM sources as a function of their angular separation. The
coincidences were selected adopting an offset between NVSS/XMM
positions of R$\leq$($\epsilon_{\rm XMM}$+$\epsilon_{\rm NVSS}$),
where $\epsilon_{\rm XMM}$ and $\epsilon_{\rm NVSS}$ are the absolute
(RA,DEC) errors in the position of the XMM and NVSS sources,
respectively. Of a total 86 sources in the histogram (shadowed
area), 41 (dashed line) refer to sources with a radio flux lower than
15 mJy. The dotted line refer to the expected XMM-NVSS sources cross-identified by chance.}
\label{fig:h1} 
\end{figure}

\section{Main results} \label{results}

After applying the first three criteria, we obtained a total of 86
coincident sources. Figure \ref{fig:h1} shows the distribution of the positionally
coincident NVSS/XMM sources with Galactic latitudes
$|b|$$\leq$10$^{\circ}$, as a function of the angular separation
between the radio and X-ray positions. Positional coincidences of the 
sources peak at 1\pri2\,. Beyond this value, the numerical distribution
of positional coincidences falls significantly, producing a decreasing tail towards the 
largest angular separations. This result is a 
consequence of the fact that radio sources with the highest fluxes
have a well-determined position (around the peak) and that radio sources
with the lowest fluxes have large positional uncertainties up to $10^{\prime\prime}$
and belong to the region with less coincidences. To assess whether 
the tail is really due to faint NVSS sources, we superimposed a histogram 
of the angular separation for sources with NVSS flux density below 15 mJy on 
the numerical distribution. As can be seen, at the largest angular separations, 
the contribution of weak sources becomes significant.

Finally, we evaluated the number of spurious identifications
due to chance alignments $N_{\rm chance}$, as a function of the
identification radius $R_{\rm id}$. Obviously $N_{\rm chance}$ should be scaled out
by sky area $A_{\rm search}$ in the study and by the cross-identification area A$_{\rm id}$=$\pi$R$_{\rm id}$$^2$.
To compute how $N_{\rm chance}$ increases, we need to increase the identification radius
to a larger one: R$_2$=R$_{\rm id}$+$\Delta$R. Thus A$_{\rm id}$ is increased
in an anular region of area $\pi$(R$_2^2$-R$_{\rm id}$$^2$)= $\pi$$\Delta$R(2R$_{\rm id}$+$\Delta$R). 
Assuming ($i-$) uncorrelated NVSS and XMM-Newton source positions and ($ii-$) 
a uniform surface source density of  NVSS sources lying in the search area, the total of
expected spurious identifications can be written as

\begin{equation}
N_{\rm chance}(R_{\rm id})= \pi\,N_{\rm X}\left({N_{\rm NVSS} \over A_{\rm search}}\,\right)\Delta\,R(2R_{\rm id}\,+\,\Delta\,R), 
\end{equation} 

\noindent where $N_{\rm X}$ and $N_{\rm NVSS}$ are the total  XMM and
NVSS sources for Galactic latitude $\leq$ 10$^{\circ}$. In Fig.1,
$\Delta\,R$ and R$_{\rm id}$ correspond to the bin of the histogram
(0.8) and offset (x-axis), respectively. We chose the cut in the identification radius 
as the largest (R$_{\rm id}$$<$10 arcsec) for which N$_{\rm true}$ $>$ N$_{\rm chance}$. As a result, we find that at
most 12 coincident sources could be expected by chance in our sample.
This value represents about 13\% of the matches. In Fig.1 we overplot a
curve with the number of coincidences expected by
chance as a function of angular separation.

As a futher step, we inspected the SIMBAD and NED database and found that, out of 86
positional coincident sources 24, are well-known objects. These include 
10 galaxies  \cite{nakanishi97}, the cluster of
galaxies ClG 0745-19 \cite{taylor94}, the globular cluster NGC 6440 
\cite{criscienzo06}, 3 stars (HD 108, Naze et al. 2004; V*V684 Mon, Rouser et al. 1988; $\zeta$ Puppis, Blomme et al. 2003), 
3 star forming regions \cite{gondoin06}, 2 pulsars (PSR J0737-3039, Chatterjee et al. 2005; PSR J1833-1034, Camilo et al. 2006), the HII region 
G10.62-0.38 (Hofner \& Churchwell 1996), and the remaining 62 objects are unidentified. 
Out of 15 well-known MQs, 5 have been detected in the NVSS survey, 3
belong to the southern sky and therefore lie outside the survey limit, and
the remaining 7 were not detected. None of these 5 radio detections is covered by the 
{\it XMM-Newton} catalog, which is why we did not recover them in this work. 

\subsection{X-ray colors of NVSS/XMM objects}


Statistical studies of high-energy sources based on X-ray colors can
be used to classify objects with different spectral-energy
distributions belonging to  different galactic populations. 
Thanks to the wider (0.2-12 keV) energy range of the SXSSC, we are
able to compute X-ray colors of sources in three different
broad-bands, thus helping us better unmask signs of highly
energetic processes in our list of positionally coincident NVSS/XMM
sources. The resulting properties of the newly-found MQ candidates are
presented in the table and plots discussed below.

We use three bands here defined in the catalog: Soft (S: 0.2-1.0 keV), 
Medium (M: 1.0-2.0 keV), and Hard (H: 2-12.0 keV). Figure 2 shows the count ratio of X-ray colors
$H_x=(M-S)/(S+M)$ and $H_y=(H-M)/(H+M)$. While it is difficult to classify
individual sources with confidence on the basis of X-ray color alone,
we computed and plot the predicted loci for absorbed power-law models with
a $\Gamma$ index ranging from 0 to 4 and an interstellar absorption
$N_{\rm H}$ from $10^{21}$ to 4$\times$ 10$^{22}$ cm$^{-2}$. The grid was
calculated with the {\it Portable Interactive Multi-Mission Simulator}
(PIMMS\footnote{http://heasarc.gsfc.nasa.gov/docs/software/tools/pimms.html})
by using a power-law emission model. 


\begin{figure*}[t!]
\resizebox{1.0\hsize}{!}{\includegraphics[angle=0]{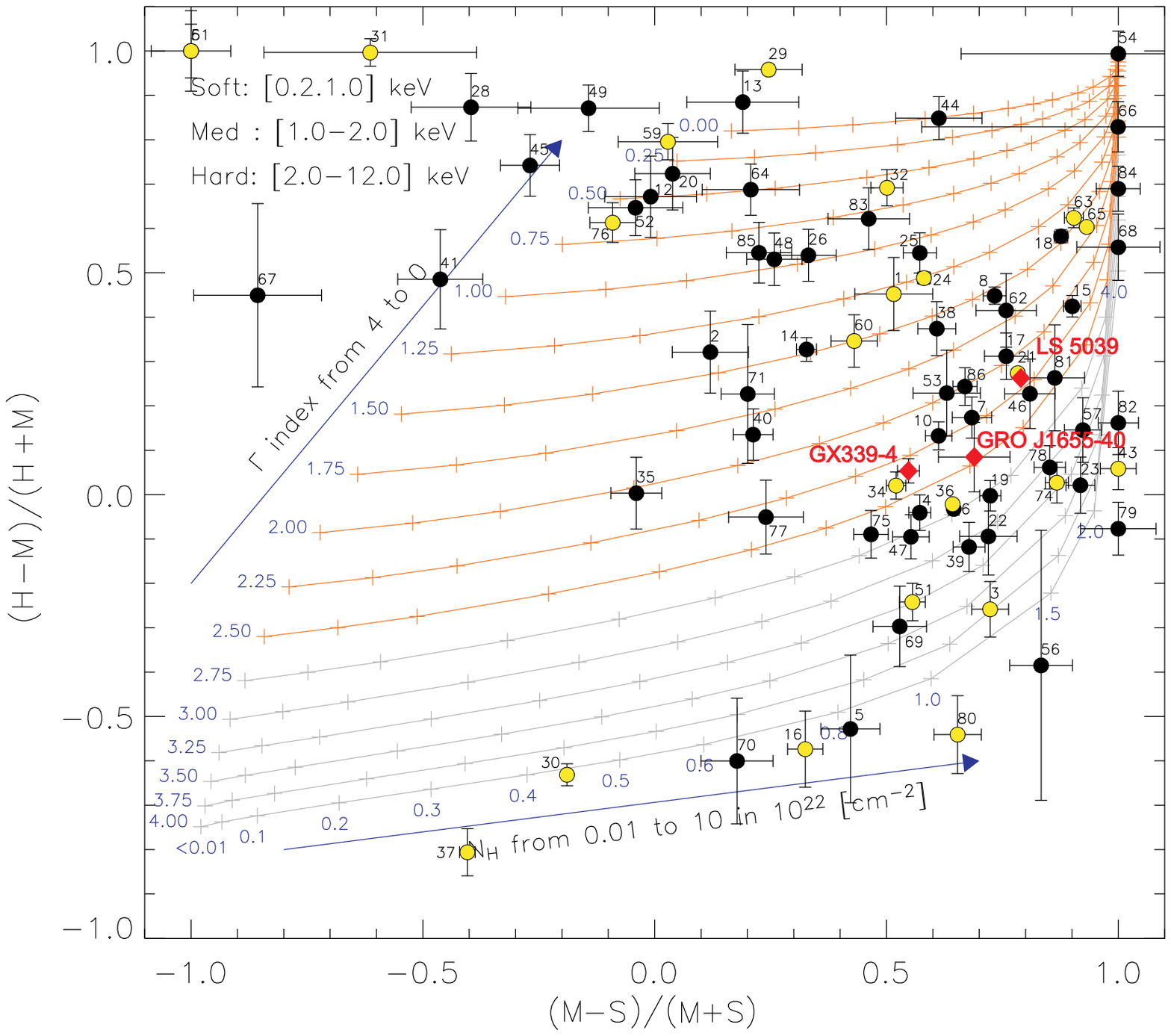}}
\caption{Ratios of source counts in different spectral bands. Soft
($S_{\rm x}$: 0.5-2.0 keV), Medium ($M_{\rm x}$: 2.0-4.5 keV), and Hard
($H_{\rm x}$: 4.5-12.0 keV). $(S-M)/(S+M)$ on $X$-axis and
$(H-M)/(H+M)$ on $Y$-axis. The $1 \sigma$ error bars are shown for all
the sources. Filled (black) points correspond to unidentified objects
in our sample while those shown as open (yellow) points correspond to
already known sources. Well-known microquasars are also indicated as (red) 
diamonds.} \label{fig:co} \end{figure*}

From a direct comparison of the source position with respect to the
$\Gamma$-$N_{\rm H}$ grid, we obtained an estimation for the $N_{\rm H}$
and $\Gamma$ values, which is a good approximation of the X-ray energy
distribution of the sources. Table 1 shows the X-ray properties of the
coincident NVSS-XMM sources. The present knowledge of each one is
labeled according to the flag-Id numbers in the  last column of the table. 
Figure 2 shows the ratios of source  counts in different spectral
bands. In this image three microquasars (LS 5039, GRO 1655-40, and
GX339-4) that are undetected in the NVSS survey or not strictly coincident with their 
SXSSC counterparts are indicated for comparison purposes. 
The position of the three MQs in the figure 
agree very well with the values of $\Gamma$ and $N_{\rm H}$ obtained by 
several authors using {\it XMM-Newton} data; e.g., Bosch-Ramon et al. (2007) 
in LS 5039; Miller et al. (2004) for GX339-4; and Diaz Trigo et al. (2007) 
for GRO J1655-40. For sources 9, 11, 27, 28, 29, 31, 33, 37, 42, 45, 49, 
50, 54, 55, 56, 58, 61, 72, 73, 79, 80, 82, and 84 that lie
outside the $N_{\rm H}$-$\Gamma$ grid, the power-law model does not
provide an adequate fit. 

In order to distinguish different types of objects and to check 
the reliability of our analysis, we compared the values 
of $N_{\rm H}$ and $\Gamma$ of the sources with the classification scheme 
presented by Prestwich et al. (2003). For this purpose, 
we applied the hardness ratio classification using the 
band definition by Prestwich et al. (2003) exactly. We computed $HR_{\rm
soft}=(M_{\rm x}-S{\rm x})/T_{\rm x}$ and $HR_{\rm hard}=(H-M)/T$,
where $S_{\rm x}$:0.3-1., $M_{\rm x}$:1-2, $H_{\rm x}$:2-8 keV, and
$T_{\rm x}$ are the total counts in all three bands combined. 

\begin{figure*}[t!] 
\resizebox{1.0\hsize}{!}{\includegraphics[angle=0]{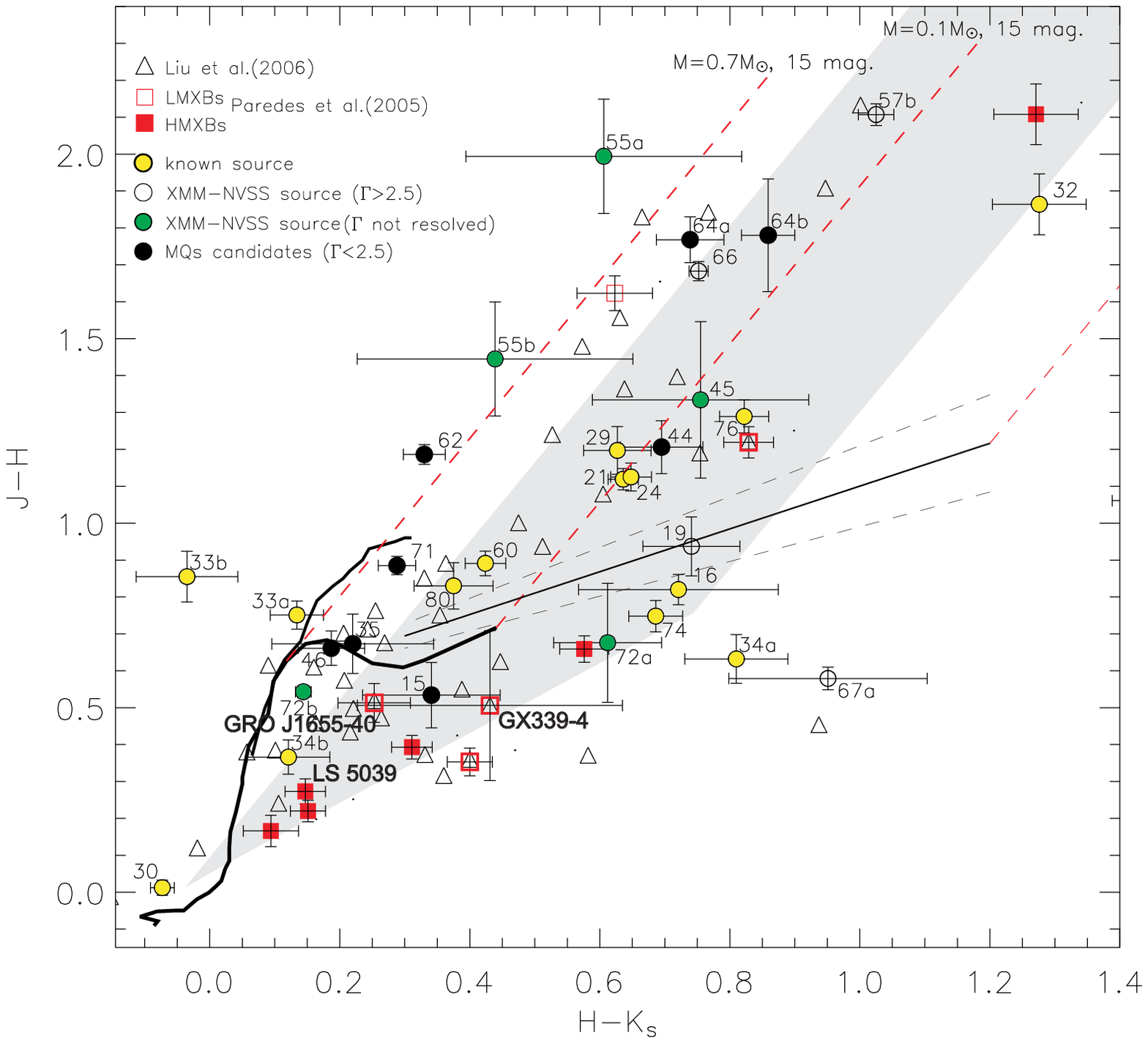}}
\caption{Color-color diagram $(J-H/H-K)$ of the positional correlated sources. 
The intrinsic colors of the main sequence \cite{siess00}, the CTTS locus \cite{meyer97}, and giant stars are shown, 
as well as the slope and magnitude of the interstellar reddening vector (A$_{\rm V}$=15 mag). Well-known MQs (in red) and
HMXB systems (white open squares) are also shown. The most promising MQ
candidates (indicated as black points) lie inside the region shaded in grey.} \label{fig:integra} \end{figure*}

We find that sources with $\Gamma<1$ and $1<\Gamma<2.5$, with a small
dependence on the $N_{\rm H}$, lie in the HMXB and LMXB loci,
respectively. This interpretation agrees with being HMXBs 
known to display spectra in the 1-10 keV region, well-represented by a
power-law index $\Gamma$ of 1\,-\,2 \cite{white95}, and also often
with a high variable intrinsic absorption to reach $\Gamma$ values as
low as $\sim$0.3. We have 39 NVSS-XMM sources in this regime, with 
only 10 of them well-known objects (1, 21, 24, 32, 34, 59, 60, 63, 65 and 76). 
The remaining 29 objects (2, 7, 8, 10, 12, 13, 14, 15, 17, 18, 20, 25, 26, 
35, 38, 40, 44, 46, 48, 52, 53, 62, 64, 71, 77, 81, 83, 85, 86) could be low-mass or
high-mass XRB systems. As expected, LS 5039, GX339-4, and GRO J1655-40 lie 
within this zone.

Since it is incomplete to define MQ candidates based on the individual
X-ray colors alone, a more accurate perspective on the near-IR
properties of the 86 NVSS-XMM objects can help us improve our
knowledge about the nature of the sources, as well as to concentrate
efforts on those objects with more reliable MQ characteristics.   

\subsection{Cross-identification of NVSS/XMM with the 2MASS near-IR data}

Given that 2MASS photometry is only complete to $K_s$ band $\approx$
14.8 mag, the cross-correlation between our NVSS-XMM source list with
the 2MASS database is very restrictive. However, the availability of
near-IR observations of the whole sky provides an excellent opportunity to
focus on our MQ candidate sample. The cross-identification of the 86 
NVSS/XMM sources with the 2MASS catalog was performed using a search
radius $R\sim 4^{\prime\prime}$ (the 90\% uncertainty radius in the X-ray position). 
In addition, one additional restrictive criteria was applied. Accretion disks with 
different ranges of accretion rates and
viewing angles account for intrinsic near-infrared excesses
distributed in a narrow range in the $(J-H)$ vs $(H-K_s)$ diagram
\cite{meyer97}. We also use this information to constraint the
location of MQ candidates based on their near-IR properties.

The resulting sample now contains 31 sources with one or two 2MASS counterpart candidates. Of these, 
16 are already cataloged sources (see flag-Id in Table 1), while the remaining 15 objects appear 
as unidentified. The fact that some of the candidates show no IR counterpart in the 2MASS catalog does not 
mean that no IR counterpart exists. They could simply be highly obscured objects. Hereafter,
we only focus our attention on these coincident NVSS-XMM sources with 2MASS counterparts. A deeper search for 
IR counterparts for the rest of the sources undetected in the 2MASS counterpart is in progress.

\subsection{Near infrared color-color analysis}

From the near-IR photometric measurements, we can locate objects in the 
$(J-H)$ vs $(H-K_s)$ color-color diagram. This allows us to distinguish objects with small, 
intermediate, or large near-infrared excess, as well as those with intrinsic $K_s$ 
excess that are probably produced by the presence of a disk-star structure \cite{meyer97}. 

Figure 3 shows the color-color diagram $(J-H)$-$(H-K_s)$ for the sources. To facilitate 
the identification in the diagram, we used for each source the number assigned in Table 
1 and Figure 2. We also plot the near-IR colors of known HMXB systems 
\cite{liu06}, well-known MQs \cite{paredes05}, the classical T-Tauri stars (CTTS) locus of Meyer et al. (1997), and 
two reddening vectors with lengths corresponding to $A_{\rm V}$=15 mag. It is interesting to note that, while MQs in HMXB systems 
have an approximate lineal behavior in the diagram, LMXBs have a more 
random distribution. 

This suggests that the most promising MQ candidates in HMXB systems
could follow the HMXB loci, which might be placed following the direction of reddening vectors
because of the different values of the interstellar absorption 
We indicate this region of possible MQ candidates with a grey 
region in Fig\,3. Six sources (36, 49, 50, 51, 57, and 59) with infrared counterparts fall outside the limits of the diagram. 
Of these, 4 are well-known objects (see flag-Id in Table 1) and 2 of them do not fulfill the characteristics required  
to be low-mass or high-mass X-ray systems; therefore, they were discarded from our analysis. As can be seen, only 
three sources (15, 44, and 64b) lie within the suggested MQ grey region. 

Finally, it is worth noting that, in the hardness ratio classification by \cite{prestwich03}, the sources 15, 44, and 64  
show values of $HR_{\rm soft}$ 0.27, 0.06, and 0.05 and $HR_{\rm hard}$ 0.42, 0.83, and 0.60, respectively. Therefore, 
these sources lie in the loci of HMXBs.

\section{Summary} \label{summary}

In this work we have presented an improved search strategy for finding new MQ candidates based on positional 
cross-identifications. By analyzing radio and modern more sensitive X-ray data, we found 86 positional-coincidence sources 
in the NVSS and XMM catalog, with galactic latitudes $|b| \leq 10^{\circ}$. 
Among them, 24 are well-known objects and the rest of the 62 sources are unidentified. Out of the 86 
sources 31 have one or two infrared counterpart candidates in the 2MASS catalog.
For all the sources, when possible, the hardness ratio was used to assess the 
low-mass or high-mass XRB properties of the objects. At least 29 unidentified objects fulfill these characteristics.  
Using the 2MASS counterpart of well-known MQs, we found that MQs with a high-mass nature follow a lineal 
behavior in the infrared color-color diagram, while LMXB MQs display a more random distribution. 
Based on this information, we suggest a possible region in the infrared diagram for new MQ candidates. 
As a result, we found 3 objects that we propose as likely galactic HMXBs and MQ candidates. 
A follow-up study of these three sources will be reported on a future paper (Combi et al. 2007). 

Finally, it is important to mention that the whole analysis carried out here only covers 
a minute fraction of the expected MQ candidates. The NVSS survey does not cover all of the sky. 
It is complete up to $-40^{\circ}$ in declination. In addition, the current XMM catalog only 
represents $\sim$ 2\% of the galactic plane ($|$b$|$ $\leq$ 10$^{\circ}$). Thus, a naive extrapolation
of the 3 MQ candidates to the whole Galactic plane suggests that about 120 new candidates could be
expected in the future.  
  
\begin{acknowledgements} 
We thank the anonymous referee for her/his insightful comments 
and constructive suggestions that lead to a substantial improvement of the manuscript. 
We also thank Valenti Bosch-Ram\'on and Marc Rib\'o for useful discussions, and 
the staff of the Osservatorio Astronomico di Palermo where part of this research 
was carried out. J.A.C. is a researcher of the programe {\em Ram\'on y
Cajal} funded jointly by the Spanish Ministerio de Educaci\'on y Ciencia (former
Ministerio de Ciencia y Tecnolog\'{\i}a) and Universidad de Ja\'en. J.F.A.C. is
a researcher of the Consejo Nacional de Investigaciones Cient\'ificas y
Tecnol\'ogicas (CONICET), Argentina. The authors acknowledge support by the DGI
of the Spanish Ministerio de Educaci\'on y Ciencia under grants
AYA2004-07171-C02-02, FEDER  funds and Plan Andaluz de Investigaci\'on of Junta
de Andaluc\'{\i}a as research group FQM-322. 

\end{acknowledgements}



\begin{figure*}[t!]
\resizebox{1.0\hsize}{!}{\includegraphics[angle=0]{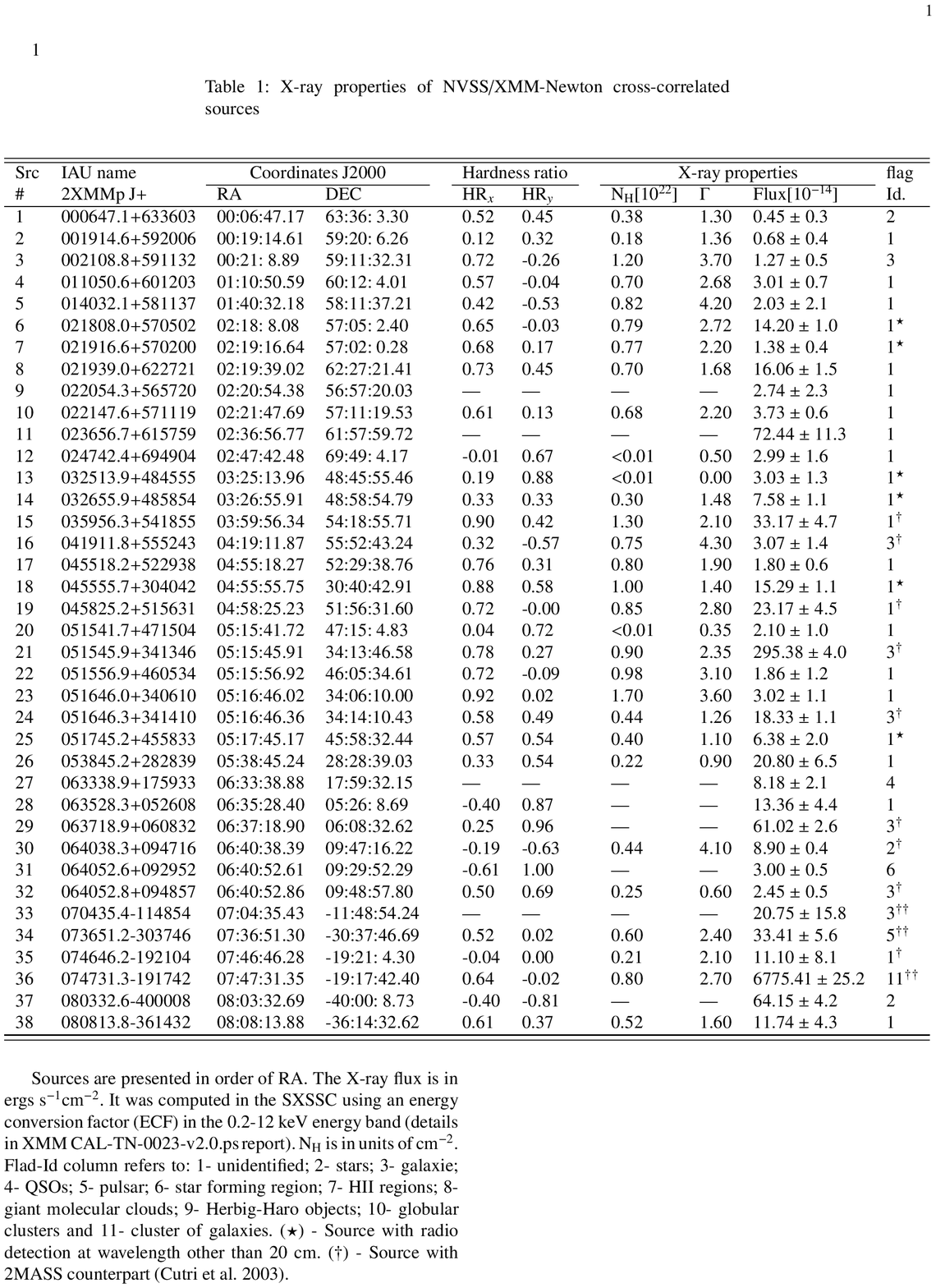}}
\caption{} \label{fig:co} \end{figure*}


\begin{figure*}[t!]
\resizebox{1.0\hsize}{!}{\includegraphics[angle=0]{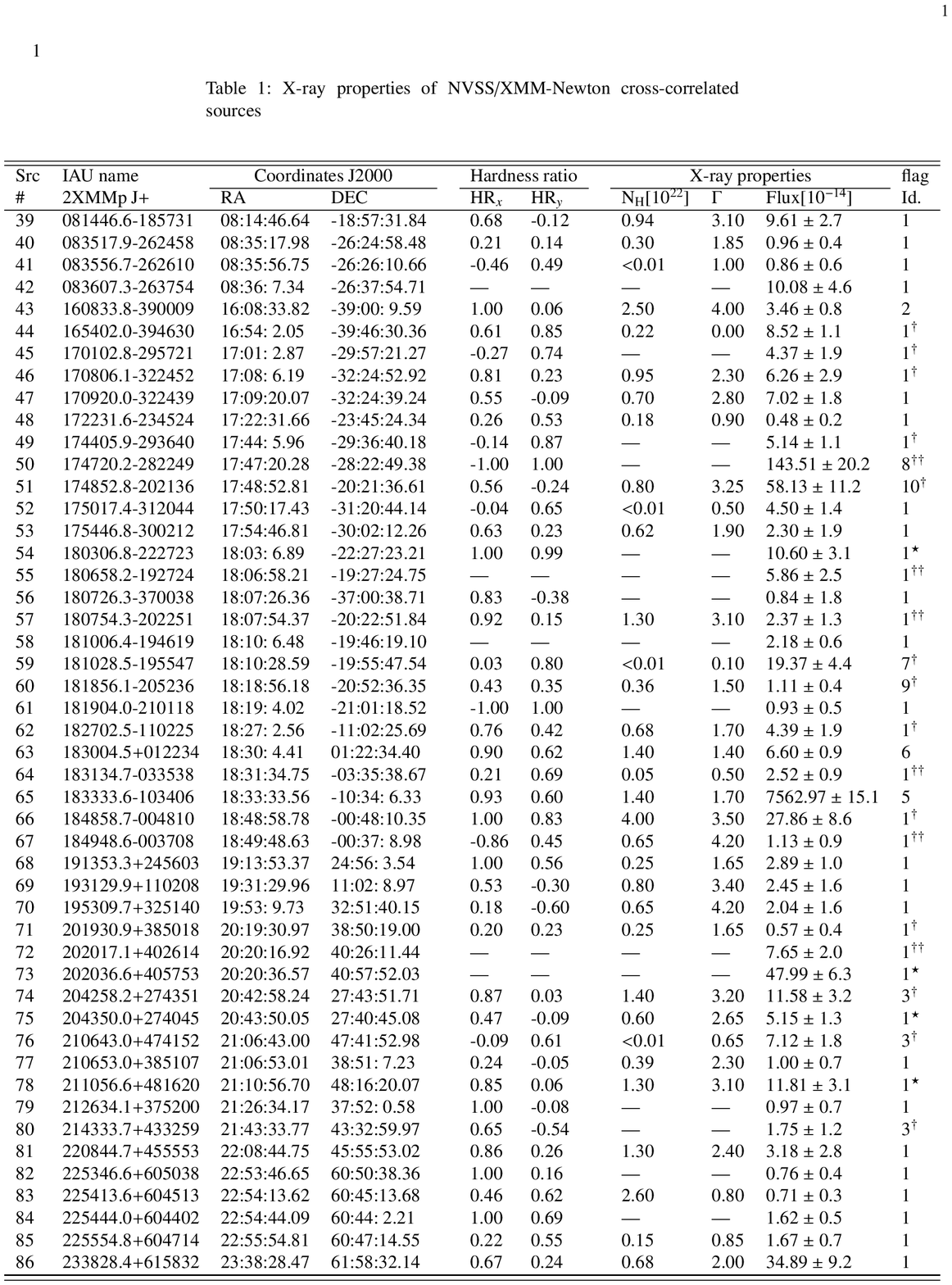}}
\caption{} \label{fig:co} \end{figure*}

\end{document}